\documentclass[aip,
amsmath,amssymb,
reprint,%
]{revtex4-1}

\usepackage{epsfig,graphicx,times}
\usepackage{textcomp}
\usepackage{amsmath,amssymb,amstext}            
\usepackage{graphicx}           
\usepackage{latexsym}
\usepackage{bm}
\usepackage{appendix}
\usepackage{amsmath}
\usepackage[usenames,dvipsnames]{color}
\usepackage{ulem}
\usepackage{float}
\usepackage{braket}
\usepackage[english]{babel}
\usepackage[colorlinks,bookmarks=false,citecolor=blue,linkcolor=red,urlcolor=blue]{hyperref}

\begin{document}
	
	\title{Evading thermal population influence on enantiomeric-specific state transfer based on a cyclic three-level system via ro-vibrational transitions }
	
	\author{Quansheng Zhang}
	\affiliation{Beijing Computational Science Research Center, Beijing 100193, China}
	
	\author{Yu-Yuan Chen}
	\affiliation{Beijing Computational Science Research Center, Beijing 100193, China}
	
    \author{Chong Ye}
    \email{yechong@bit.edu.cn}
    \affiliation{Beijing Key Laboratory of Nanophotonics and Ultrafine Optoelectronic Systems,
    	School of Physics, Beijing Institute of Technology, 100081 Beijing, China}
	\affiliation{Beijing Computational Science Research Center, Beijing 100193, China}

	\author{Yong Li}
	\email{liyong@csrc.ac.cn}
	\affiliation{Beijing Computational Science Research Center, Beijing 100193, China}
	\affiliation{Synergetic Innovation Center for Quantum Effects and Applications, Hunan Normal University, Changsha 410081, China}
	\date{\today}

\begin{abstract}
Optical methods of enantiomeric-specific state transfer had been proposed theoretically
based on a cyclic three-level system of chiral molecule. According to these theoretical methods, recently the breakthrough progress has been reported in experiments {[}S. Eibenberger \textit{et al.}, Phys. Rev. Lett. \textbf{118}, 123002 (2017);  C. P\'{e}rez \textit{et al.}, Angew. Chem. Int. Ed. \textbf{56}, 12512 (2017){]} for cold gaseous chiral molecules but with {achieving} low state-specific enantiomeric enrichment. One of the limiting factors is the influence of the thermal population {in the selected cyclic three-level system based on purely rotational transitions} in experiment. Here, we theoretically explore the improvement of the enantiomeric-specific state transfer at finite temperature by introducing ro-vibrational transitions for the cyclic three-level system of chiral molecules. Then, at the typical temperature in experiments, approximately only the lowest state in the {desired cyclic} three-level system is thermally occupied and the optical method of enantiomeric-specific state transfer works well. Comparing with the case of purely rotational transitions where all the three states are thermally occupied, this modification will remarkably increase the obtained state-specific enantiomeric enrichment with enantiomeric excess {being} approximately 100\%.
\end{abstract}
\maketitle

\section{Introduction}\label{Introduction}

Molecular chirality has played a crucial role in the enantio-selective biological and chemical processes~\cite{Book_UMeierhenrich}, homo-chirality of life~\cite{Satio_Chiral_Review_RevModPhys.85.603},
and even fundamental physics~\cite{MQuck_PVED_Angew1989}.  It has attracted considerable interests to realize enantioseparation and enantiodiscrimination not only in chemistry~\cite{Book_ChiralAnalysis} but also in atomic, molecular, and optical physics~\cite{CY_Con1,CY_Con2,Shapiro_Con,EFishman_Con,Ye_PRR_Con,PBrumer_Con,PKral_PRL_Con,Xia,WZJ_Detection,Patterson_Detection,DPatterson_Detection2,Patterson_Detection3,Patterson_Detection4,Patterson_Detection5,Buhmann_Detection1,Buhmann_Sep,Koch_Detec,KKLehmann_arXiv,YY_Chen,CY_ACStark,Lobsiger,Shubert,Delta,Shapiro_PopulationTransfer_PhysRevLett.87.183002,Shapiro_OpticalSwitch_PhysRevLett.90.033001,YLi_PauseScheme_PhysRevA.77.015403,Jia_2010_OpticalPause,Effective_two_level_Model,Vitan_Shortcut_PhysRevLett.122.173202,YLi_SpatialSepatation_PhysRevLett.99.130403,LiXuan_SpatialSeparation,DPatteson_exp_PhysRevLett.118.123002,Exp_Coherent_Enantiomer_Selective,QuantumControl,Koch_Principles_JCP,Diastereomer,Exp_supersonic,Exp_Cooling_Supersonic}. Among these, solely optical (or microwave) methods with the framework of cyclic three-level ($\Delta$-type) system (CTLS)~\cite{Liu_CyclicThreeLevel,YL_CQED,Jia_CQED} based on electronic-dipole transitions, have caught the attention to realize enantioseparation~\cite{Xia,Shapiro_PopulationTransfer_PhysRevLett.87.183002,Shapiro_OpticalSwitch_PhysRevLett.90.033001,YLi_PauseScheme_PhysRevA.77.015403,Jia_2010_OpticalPause,Effective_two_level_Model,Vitan_Shortcut_PhysRevLett.122.173202,YLi_SpatialSepatation_PhysRevLett.99.130403,LiXuan_SpatialSeparation,DPatteson_exp_PhysRevLett.118.123002,Exp_Coherent_Enantiomer_Selective,QuantumControl,Koch_Principles_JCP,Exp_supersonic,Exp_Cooling_Supersonic} as well as enantiodiscrimination~\cite{WZJ_Detection,Patterson_Detection,DPatterson_Detection2,Patterson_Detection3,YY_Chen,CY_ACStark,Lobsiger,Shubert,Delta,Patterson_Detection4,Patterson_Detection5,KKLehmann_arXiv,Grabow}
\textcolor{black}{and enantioconversion~\cite{CY_Con1,CY_Con2,Shapiro_Con,EFishman_Con,Ye_PRR_Con,PBrumer_Con,PKral_PRL_Con}}.

\textcolor{black}{The CTLS of chiral molecules is special since the electric-dipole transition moments for the three transitions are proportional to the projections of the electric-dipole transition moment onto the three inertial axes {of gaseous chiral molecules}, respectively. It {is} well known that their product changes sign for the enantiomers \cite{Y_RealSingleLoop_PhysRevA.98.063401,Grabow,DPatterson_Detection2}. This inherent symmetry of the system can be mapped on the coupling strengths in the CTLS of chiral molecules~\cite{Delta,Shapiro_PopulationTransfer_PhysRevLett.87.183002,Shapiro_OpticalSwitch_PhysRevLett.90.033001,YLi_PauseScheme_PhysRevA.77.015403,Jia_2010_OpticalPause,Effective_two_level_Model,Vitan_Shortcut_PhysRevLett.122.173202,YLi_SpatialSepatation_PhysRevLett.99.130403,LiXuan_SpatialSeparation,DPatteson_exp_PhysRevLett.118.123002,Exp_Coherent_Enantiomer_Selective,QuantumControl,WZJ_Detection,Patterson_Detection,DPatterson_Detection2,Patterson_Detection3,YY_Chen,CY_ACStark,KKLehmann,KKLehmann2,Koch_Principles_JCP,Lobsiger,Shubert,Delta,Exp_supersonic,Exp_Cooling_Supersonic,Patterson_Detection4,Patterson_Detection5,KKLehmann_arXiv,Xia,Grabow}, i.e., the product of the three coupling strengths changes sign for the enantiomers.}
Using this feature, one can achieve the enantiomeric-specific state transfer~\cite{DPatteson_exp_PhysRevLett.118.123002,Exp_Coherent_Enantiomer_Selective,Exp_supersonic,Xia,Exp_Cooling_Supersonic}
(also named as inner-state enantioseparation~\cite{Effective_two_level_Model,YLi_PauseScheme_PhysRevA.77.015403,Jia_2010_OpticalPause,Effective_two_level_Model}), {i.e., for an initial racemic chiral mixture, finally the populations in certain energy levels (e.g. the ground states) will be different for the enantiomers.} Based on the CTLS via optical methods such as {adiabatical processes}~\cite{Shapiro_PopulationTransfer_PhysRevLett.87.183002,Shapiro_OpticalSwitch_PhysRevLett.90.033001},
a dynamic operation of ultrashort pulses~\cite{YLi_PauseScheme_PhysRevA.77.015403,Jia_2010_OpticalPause,Effective_two_level_Model}, and shortcuts-to-adiabaticity operations~\cite{Vitan_Shortcut_PhysRevLett.122.173202,QuantumControl}, in principle the perfect  enantiomeric-specific state transfer with {100\% enantiomeric excess} can be achieved with {the two enantiomers} occupying different-energy levels. Furthermore, the left- and right-handed molecules in {different-energy levels} can be spatially separated by a variety of energy-dependent \textcolor{black}{processes}~\cite{Shapiro_OpticalSwitch_PhysRevLett.90.033001,Shapiro_PopulationTransfer_PhysRevLett.87.183002,Vitan_Shortcut_PhysRevLett.122.173202}.
In addition, based on the similar CTLS, one can also achieve directly the spatial enantioseparation by mean of the chiral generalized Stern-Gerlach effect~\cite{YLi_SpatialSepatation_PhysRevLett.99.130403,LiXuan_SpatialSeparation}, that is, {the two enantiomers} move along different spatial trajectories.

Recently, breakthrough experiments~\cite{DPatteson_exp_PhysRevLett.118.123002,Exp_Coherent_Enantiomer_Selective,
Exp_supersonic,Exp_Cooling_Supersonic} have reported the enantiomeric-specific state transfer for gaseous chiral molecules with the similar ideas proposed in Refs.~\cite{YLi_PauseScheme_PhysRevA.77.015403,Jia_2010_OpticalPause}. However, the obtained enantiomeric enrichment is {only about 6\%}~\cite{DPatteson_exp_PhysRevLett.118.123002,Exp_Coherent_Enantiomer_Selective,Exp_supersonic,Exp_Cooling_Supersonic}. One of the important factors limiting the obtained enantiomeric enrichment is that {each of the three selected rotational states in the CTLS may have multiple degenerated magnetic sub-levels. That means the considered CTLS will be of multi loops. Thus,} the ability of enantiomeric-specific state transfer will be suppressed~\cite{DPatteson_exp_PhysRevLett.118.123002,Horberger_rotation}. Later on, it was pointed out in the theoretical work~\cite{Y_RealSingleLoop_PhysRevA.98.063401,Koch_Principles_JCP} {that} the real single-loop CTLS of \textcolor{black}{gaseous chiral molecules of asymmetric top} can be constructed by appropriately choosing three optical (microwave) fields. Another important factor limiting the obtained enantiomeric enrichment is the thermal population on the three levels of CTLS. In those experiments~\cite{DPatteson_exp_PhysRevLett.118.123002,Exp_Coherent_Enantiomer_Selective,Exp_supersonic,Exp_Cooling_Supersonic}, the CTLS consists of only rotational transitions, whose transition frequencies are usually at microwave wavelengths. Thus, the populations in the three levels will approximately have the same order of magnitude at the typical effective rotational temperature  $\sim$10\,K~\cite{DPatteson_exp_PhysRevLett.118.123002,Exp_Coherent_Enantiomer_Selective,Exp_supersonic,Exp_Cooling_Supersonic}
according to the Boltzmann distribution. This will bring an adverse {influence} on {the efficiency of enantiomeric-specific state transfer based on the CTLS}.

\textcolor{black}{In this paper, we aim to evade the adverse {influence} of the thermal population on the enantiomeric-specific state transfer. {For this purpose, we re-construct the desired CTLS of (gaseous) chiral molecules by choosing the ground state and other two excited states to have different vibrational sub-levels. Then the transition frequencies between the ground state of the desired CTLS and the two excited states are much larger than that of the previous ones as used in the experiments~\cite{DPatteson_exp_PhysRevLett.118.123002,Exp_Coherent_Enantiomer_Selective,	Exp_supersonic,Exp_Cooling_Supersonic}, where all the three states have the same vibrational sub-level. Thus, in the desired CTLS, approximately only the ground state is occupied and the adverse {influence} of the thermal population on the enantiomeric-specific state transfer can be evaded. We use the 1,2-propanediol as an example to demonstrate our idea. Specifically, the vibrational sub-level of the ground state (the two excited states) in the desired CTLS is chosen to be the ground state (first excited state) of the vibrational degree of freedom corresponding to OH-stretch.} 
The transition frequency between the two chosen vibrational {sub-levels is about $2\pi \times 100.9500$\,THz~\cite{Propanediol_Vib}. It is indeed that} almost only the ground state in the desired CTLS is thermally occupied initially in a wide range of temperature ($0-300$\,K). Thus, the {influence} of the thermal population on the enantiomeric-specific state transfer is evaded.}

The structure of this paper is organized as following. In Sec.~\ref{EnantiomerTransfer},
we describe the optical method of the \textcolor{black}{enantiomeric-specific state  transfer}
based on a dynamic operation of ultrashort pulses. In Sec.~\ref{ro-vibrationalStates},
we investigate the influence of finite temperature on the
\textcolor{black}{enantiomeric-specific state  transfer}, and compare the obtained state-specific enantiomeric
enrichment via ro-vibrational transitions with that in the same vibrational case {via purely rotational transitions}. Finally, we summarize the conclusions in Sec.~\ref{Conclution}.

\section{dynamic operation of ultrashort pulses}\label{EnantiomerTransfer}

\begin{figure}
\centering{}\includegraphics[scale=0.4]{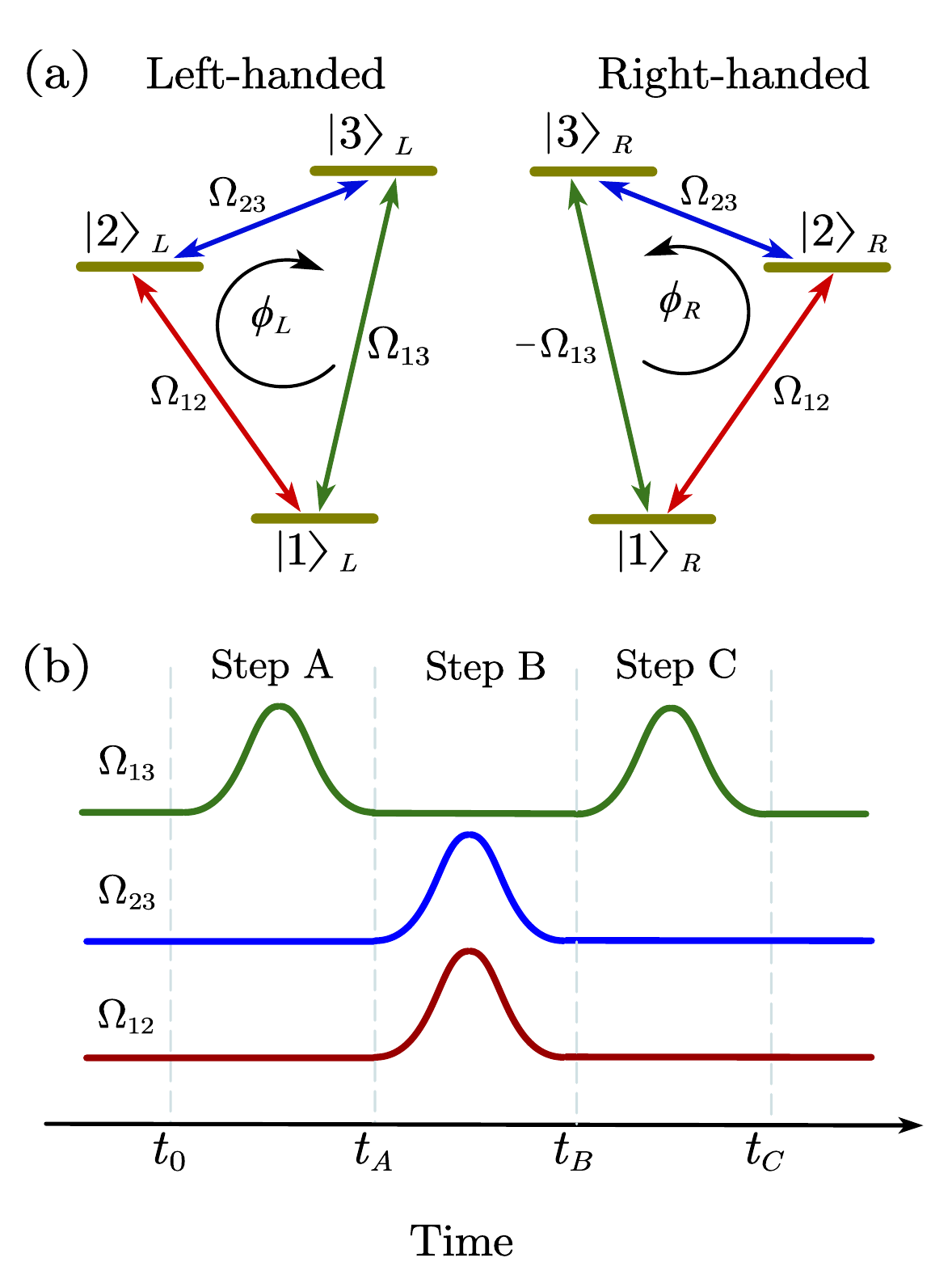} \caption{(Color online) (a) Model of CTLSs for left- and right-handed molecules.
The three-level system is resonantly coupled to the three classical
fields with coupling strengths $\pm\Omega_{13}$, $\Omega_{12}$,
and $\Omega_{23}$, respectively. \textcolor{black}{{The overall phase is depicted along the circle arrow with $ \phi_L $ and $ \phi_R $ for left- and right-handed molecules, respectively. They satisfy $\phi_R=\phi_L+\pi$}.}  (b) Schematic representation of ultrashort optical pulses to achieve perfect \textcolor{black}{enantiomeric-specific state  transfer} with the molecule initially prepared in the ground state, e.g. $\vert1\rangle_{Q}$. The related coupling strengths satisfy $\int_{t_{0}}^{t_{A}}\Omega_{13}(t)\ dt=\pi/4=\int_{t_{A}}^{t_{B}}\Omega_{0}(t)\ dt/2=-\int_{t_{B}}^{t_{C}}\Omega_{13}(t)\ dt$
with $\Omega_{23}(t)=\vert\Omega_{23}(t)\vert=-i\Omega_{12}(t)=\Omega_{0}(t)/\sqrt{2}$.
\label{fig:Structure-Pulse}}
\end{figure}

We now consider the CTLS of a chiral molecule coupled with three
classical electromagnetic (optical or microwave) fields as shown in
Fig.~\ref{fig:Structure-Pulse}(a). Then the Hamiltonian for the
system in the interaction picture with respect to $H_{0}^{Q}=\sum_{n}\hbar\omega_{n}\vert n\rangle_{Q}{}_{Q}\langle n\vert$
reads
\begin{equation}
H^{Q}(t)=\sum_{m>n=1}^{3}\hbar\Omega_{nm}^{Q}(t)e^{i\Delta_{mn}t}\vert n\rangle_{Q}{}_{Q}\langle m\vert+\mathrm{H.c}.,\label{eq:General_Hamiltonian}
\end{equation}
where $\hbar\omega_{n}$ is the eigen-energy of state (level) $\vert n\rangle_Q$
with $Q=L$ ($Q=R$) denoting the left-handed (right-handed) chiral
molecule. The detuning for the  transition $\left\vert m\right\rangle _{Q}\rightarrow\left\vert n\right\rangle _{Q}$ is defined
as $\Delta_{mn}=\nu_{mn}-\omega_{m}+\omega_{n}$, where $\nu_{mn}$
is the frequency of the classical electromagnetic field coupling to
the corresponding transition
with the coupling strength (also called Rabi frequency) $\Omega_{mn}^{Q}(t)$. We can specify the chirality-dependence of our model via choosing the coupling strengths of the left- and right-handed molecules as~\cite{Shapiro_OpticalSwitch_PhysRevLett.90.033001,Shapiro_PopulationTransfer_PhysRevLett.87.183002,CY_ACStark,YLi_PauseScheme_PhysRevA.77.015403} $\Omega_{mn}^{L}(t)=\Omega_{mn}(t)$, $\Omega_{13}^{R}(t)=-\Omega_{13}(t)$, $\Omega_{12}^{R}(t)=\Omega_{12}(t)$, and $\Omega_{23}^{R}(t)=\Omega_{23}(t)$. \textcolor{black}{
That is, the overall phase in the CTLS differs with $\pi$ for the two enantiomers: $\phi_R = \phi_L + \pi$, as seen in Fig.~\ref{fig:Structure-Pulse}(a).}

In the following, we will briefly give the protocol of dynamic operation of ultrashort pulses~\cite{YLi_PauseScheme_PhysRevA.77.015403} for the enantiomeric-specific
state transfer, which is very similar to that of current experiments~\cite{DPatteson_exp_PhysRevLett.118.123002,Exp_Coherent_Enantiomer_Selective,Exp_supersonic,
Exp_Cooling_Supersonic}. As depicted in Fig.~\ref{fig:Structure-Pulse}(b), this protocol consists of three steps. In each step, the transitions of interest $ \ket{m}_Q\rightarrow\ket{n}_Q $ are driven resonantly, i.e. $\Delta_{mn}=0$.

In step A, only the pump pulse $\Omega_{13}(t)$ is turned on.
The corresponding Hamiltonian reads $H_{A}^{Q}(t)=\hbar\Omega_{13}^{Q}(t)(\vert1\rangle_{Q}{}_{Q}\langle3\vert+\mathrm{H.c.)}$ by assuming  coupling strength $ \Omega_{13}^{Q}$.
By controlling the pulse to make it satisfy $\int_{t_{0}}^{t_{A}}\Omega_{13}(t)\,dt=\pi/4$,
a general state (density matrix) $\rho_{0}^{Q}$ at time $t_{0}$
will evolve to $\rho_{A}^{Q}=U_{A}^{Q}\rho_{0}^{Q}U_{A}^{Q\dagger}$
at time $t_{A}$ according to the unitary evolution operator $U_{A}^{Q}=\exp(-i\int_{t_{0}}^{t_{A}}H_{A}^{Q}\,dt/\hbar)$.
In the basis \{$\vert1\rangle_{Q}$, $\vert2\rangle_{Q}$, $\vert3\rangle_{Q}$\},
the unitary evolution operators $U_{A}^{Q}$ are
\begin{equation}
U_{A}^{L}=\left(\begin{array}{ccc}
\frac{1}{\sqrt{2}} & 0 & -\frac{i}{\sqrt{2}}\\
0 & 1 & 0\\
-\frac{i}{\sqrt{2}} & 0 & \frac{1}{\sqrt{2}}
\end{array}\right),\ U_{A}^{R}=\left(\begin{array}{ccc}
\frac{1}{\sqrt{2}} & 0 & \frac{i}{\sqrt{2}}\\
0 & 1 & 0\\
\frac{i}{\sqrt{2}} & 0 & \frac{1}{\sqrt{2}}
\end{array}\right).\label{eq:Translation_Operators}
\end{equation}

In step B, we turn off the pump pulse $\Omega_{13}$ [i.e. $\Omega_{13}(t)=0$]
and introduce two ultrashort pulses $\Omega_{12}(t)$ and $\Omega_{23}(t)$
with $\Omega_{23}(t)=\vert\Omega_{23}(t)\vert=-i\Omega_{12}(t)\equiv\Omega_{0}(t)/\sqrt{2}$.
The Hamiltonian for this step reads $H_{B}^{Q}(t)=\hbar\Omega_{0}(t)(\vert D\rangle_{Q}{}_{Q}\langle2\vert+\mathrm{H.c.})$
with $\vert D\rangle_{Q}=(i\vert1\rangle_{Q}+\vert3\rangle_{Q})/\sqrt{2}$.
Under the condition $\int_{t_{A}}^{t_{B}}\Omega_{0}(t)\,dt=\pi/2$,
the state will go to $\rho_{B}^{Q}=U_{B}^{Q}\rho_{A}^{Q}U_{B}^{Q\dagger}$
with the unitary evolution operators
\begin{equation}
U_{B}^{L}=U_{B}^{R}=\left(\begin{array}{ccc}
\frac{1}{2} & \frac{1}{\sqrt{2}} & -\frac{i}{2}\\
-\frac{1}{\sqrt{2}} & 0 & -\frac{i}{\sqrt{2}}\\
\frac{i}{2} & -\frac{i}{\sqrt{2}} & \frac{1}{2}
\end{array}\right)
\end{equation}
in the basis \{$\vert1\rangle_{Q}$, $\vert2\rangle_{Q}$, $\vert3\rangle_{Q}$\}.

The step C is realized by taking $\Omega_{12}(t)=\Omega_{23}(t)=0$ and re-turning
on the pump pulse $\Omega_{13}(t)$. The pump pulse $\Omega_{13}(t)$ fulfills $\int_{t_{B}}^{t_{C}}\Omega_{13}(t)\,dt=-\pi/4$
[or equivalently $(k+3/4)\pi$ with integer $ k $]. Then, the final state reads $\rho_{C}^{Q}=U_{C}^{Q}\rho_{B}^{Q}U_{C}^{Q\dagger}$
with the unitary evolution operators
\begin{equation}
U_{C}^{L}=\left(\begin{array}{ccc}
\frac{1}{\sqrt{2}} & 0 & \frac{i}{\sqrt{2}}\\
0 & 1 & 0\\
\frac{i}{\sqrt{2}} & 0 & \frac{1}{\sqrt{2}}
\end{array}\right),U_{C}^{R}=\left(\begin{array}{ccc}
\frac{1}{\sqrt{2}} & 0 & -\frac{i}{\sqrt{2}}\\
0 & 1 & 0\\
-\frac{i}{\sqrt{2}} & 0 & \frac{1}{\sqrt{2}}
\end{array}\right).
\end{equation}

With the above three operational steps, the initial state $\rho_{0}^{Q}$
will go to the final one $\rho_{C}^{Q}=U_{Q}\rho_{0}^{Q}U_{Q}^{\dagger}$
with the total unitary evolution operator $U_{Q}=U_{C}^{Q}U_{B}^{Q}U_{A}^{Q}$
given as
\begin{equation}
U_{L}=\left(\begin{array}{ccc}
1 & 0 & 0\\
0 & 0 & -i\\
0 & -i & 0
\end{array}\right),\ U_{R}=\left(\begin{array}{ccc}
0 & 1 & 0\\
-1 & 0 & 0\\
0 & 0 & 1
\end{array}\right).\label{eq:Left_Handed_Rotation}
\end{equation}
This indicates that the two enantiomers will suffer different evolutions even when their initial states have the same form. Specially, $U_{L}$ will exchange the populations of left-handed molecules in states $\left|2\right\rangle _{L}$ and  $\left|3\right\rangle _{L}$, while $U_{R}$ will exchange the populations of right-handed molecules in states $\left|1\right\rangle _{R}$ and $\left|2\right\rangle _{R}$. When the above protocol is applied to the chiral mixture composed of chiral molecules in the two ground states $ \left|1\right\rangle _{Q} $, the perfect  enantiomeric-specific state transfer is achieved according to Eq.~(\ref{eq:Left_Handed_Rotation}), since the {initial} state $\left|1\right\rangle _{L}$ will evolve back to itself and the {initial} state $\left|1\right\rangle _{R}$ in the meanwhile will be transferred to $ \left|2\right\rangle _{R} $. After that, the left- and right-handed molecules in {different-energy} states can be further spatially separated by a variety of energy-dependent processes~\cite{Shapiro_OpticalSwitch_PhysRevLett.90.033001,Shapiro_PopulationTransfer_PhysRevLett.87.183002,Vitan_Shortcut_PhysRevLett.122.173202}.

\section{Transition Between Different Ro-vibrational States}

\label{ro-vibrationalStates}

As discussed in the above section, the perfect enantiomeric-specific state transfer can be achieved if the left- and right-handed molecules has been initially prepared in the ground state $ \ket{1}_Q $ of the CLTS. \textcolor{black}{In experiments, the effective rotational temperature of chiral mixture is usually cooled to be $\sim$10\,K with recent \textcolor{black}{technologies}, such as buffer gas cooling~\cite{DPatteson_exp_PhysRevLett.118.123002,Exp_Coherent_Enantiomer_Selective,Exp_Cooling_Supersonic} {and} supersonic expansions~\cite{Exp_supersonic,Exp_Cooling_Supersonic}.}
The initial thermal population in each state of the CTLS based on purely rotational transitions will have the same order of magnitude. This will bring an adverse {influence} \textcolor{black}{on} achieving perfect  enantiomeric-specific state transfer~\cite{DPatteson_exp_PhysRevLett.118.123002,Exp_Coherent_Enantiomer_Selective,Exp_supersonic,Exp_Cooling_Supersonic}.

Now, we investigate the thermal population influence {on} the enantiomeric-specific state transfer based on {the desired} CTLS of chiral molecules at finite temperature. For the gaseous molecules, the rotational and vibrational degrees of freedom should be taken into account~\cite{Horberger_rotation} by adopting the ro-vibrational state~\cite{Horberger_rotation,Koch_Principles_JCP}
\begin{equation}
\left\vert \psi\right\rangle =\left\vert \psi_{\mathrm{vib}}\right\rangle \vert\psi_{\mathrm{rot}}\rangle\label{state},
\end{equation}
\textcolor{black}{with the product form of the vibrational state $\vert\psi_{\mathrm{vib}}\rangle$
and the rotational state  $\vert\psi_{\mathrm{rot}}\rangle$. Here, {we have taken the rigid-rotor approximation~\cite{BooK_AngularMomentum}, which means the coupling between vibrational and rotational states under field-free conditions can be neglected. 
This assumption is available for some kinds of chiral molecules, but is not available for some others. In this paper, our discussions focus only on the case where the rigid-rotor approximation is available.}}

\subsection{Enantiomeric-specific state transfer at presence of thermal population}\label{ro-vibrationalStates_SSET}
In realistic experiments using buffer gas cooling~\cite{DPatteson_exp_PhysRevLett.118.123002,Patterson_Detection,DPatterson_Detection2,Patterson_Detection3,Patterson_Detection4,Patterson_Detection5}, the vibrational relaxation cross sections of molecule-buff gas are usually smaller than rotational ones. This indicates the effective vibrational temperature $T_{\mathrm{vib}}$ of chiral molecules is typically higher than the effective {rotational} temperature $T_{\mathrm{rot}}$ \cite{DPatterson_Cooling_MolPhys,DPatterson_Cooling_PRA,Koch_Cooling_arXiv}.
Accordingly, we assume that the effective rotational temperature to be $T_{\mathrm{rot}}\sim$10\,K and the effective vibrational temperature $ T_{\mathrm{vib}}=300$\,K.

For the three-level system, the initial thermal population in the
ro-vibrational state $\left\vert n\right\rangle _{Q}$ with the eigen-energy
$\hbar\omega_{n}=\hbar(\omega_{n,\mathrm{vib}}+\omega_{n,\mathrm{rot}})$
has the Boltzmann-distribution form
\begin{equation}
p_{n}=\frac{1}{Z}\mathcal{P}_{n,\mathrm{vib}}\mathcal{P}_{n,\mathrm{rot}}.\label{eq:Boltzmann_Distribution}
\end{equation}
Here, the factor $\mathcal{P}_{n,\mathrm{vib}}$ ($\mathcal{P}_{n,\mathrm{rot}}$)
is defined as $\mathcal{P}_{n,\mathrm{vib}}=\exp(-\hbar\omega_{n,\mathrm{vib}}/k_{B}T_{\mathrm{vib}})$
{[}$\mathcal{P}_{n,\mathrm{rot}}=\exp(-\hbar\omega_{n,\mathrm{rot}}/k_{\mathrm{B}}T_{\mathrm{rot}})${]}
with the vibrational (rotational) eigen-energy $\hbar\omega_{n,\mathrm{vib}}$
($\hbar\omega_{n,\mathrm{rot}}$), and $Z=\sum_{n=1}^{3}\mathcal{P}_{n,\mathrm{vib}}\mathcal{P}_{n,\mathrm{rot}}$
\textcolor{black}{is} the partition function. The initial state \textcolor{black}{of the CTLS for enantiomers} reads
\begin{equation}
\rho_{\mathrm{0}}^{Q}=p_{1}\left\vert 1\right\rangle _{QQ}\left\langle 1\right\vert +p_{2}\left\vert 2\right\rangle _{QQ}\left\langle 2\right\vert +p_{3}\left\vert 3\right\rangle _{QQ}\left\langle 3\right\vert,
\end{equation}
with $p_1+p_2+p_3=1$.

By means of the dynamic operation of ultrashort pulses as discussed in the above section, the states $\rho_{0}^{L,R}$ for enantiomers will evolve differently under the different evolution operators $U_{L,R}$. Then the final state for the left-handed molecules is given as
\begin{align}
\rho_{C}^{L} & =p_{1}\left\vert 1\right\rangle _{LL}\left\langle 1\right\vert +p_{3}\left\vert 2\right\rangle _{LL}\left\langle 2\right\vert +p_{2}\left\vert 3\right\rangle _{LL}\left\langle 3\right\vert ,\label{eq:Final_L}
\end{align}
and the final state for the right-handed molecules will go to
\begin{equation}
\rho_{C}^{R}=p_{2}\left\vert 1\right\rangle _{RR}\left\langle 1\right\vert +p_{1}\left\vert 2\right\rangle _{RR}\left\langle 2\right\vert +p_{3}\left\vert 3\right\rangle _{RR}\left\langle 3\right\vert .\label{eq:Final_R}
\end{equation}
By comparing Eq.~(\ref{eq:Final_L}) with Eq.~(\ref{eq:Final_R}), there are different populations in energy-degenerated states $\left\vert n\right\rangle _{L}$ and $\left\vert n\right\rangle _{R}$ $ (n=1,2,3) $. Then, the (imperfect) enantiomeric-specific state transfer has been achieved. Specially, we now focus on the populations in \textcolor{black}{the state $\left\vert 2\right\rangle $ by introducing the} enantiomeric excess~\cite{Book_ChiralAnalysis,CY_ACStark,YY_Chen}
\begin{align}
\begin{split}
\epsilon&=\left|\frac{\rho_{C,2}^L-\rho_{C,2}^R}{\rho_{C,2}^L+\rho_{C,2}^R}\right|=\left|\frac{p_3-p_1}{p_3+p_1}\right|,\\
&=\left|1-\frac{2}{1+\frac{\mathcal{P}_{1,\mathrm{vib}}}{\mathcal{P}_{3,\mathrm{vib}}}\times\frac{\mathcal{P}_{\mathrm{1,rot}}}{\mathcal{P}_{\mathrm{3,rot}}}}\right|\label{eq:Enantiomeric_Excess-1},
\end{split}
\end{align}
with $ \rho_{C,2}^{Q}={}_{Q}\langle2\vert\rho_{C}^{Q}\left\vert 2\right\rangle _{Q} $. It gives the excess of one enantiomer over the other in the state $\left\vert 2\right\rangle $. The perfect enantiomeric-specific state transfer is achieved, when the state $ \ket{2} $ is occupied by enantio-pure molecules with $\epsilon=100\% $.

For the selected three states of the CTLS with same vibrational state $(\omega_{1,\mathrm{vib}}=\omega_{2,\mathrm{vib}}=\omega_{3,\mathrm{vib}})$,
 $\epsilon$ will be only dependent on the rotational transition
frequency $(\omega_{3,\mathrm{rot}}-\omega_{1,\mathrm{rot}})$.
If we choose different vibrational states with $\omega_{1,\mathrm{vib}}\neq\omega_{3,\mathrm{vib}}$,
the ratio $(\mathcal{P}_{1,\mathrm{vib}}/\mathcal{P}_{3,\mathrm{vib}})$, which is usually much larger than 1, will bring an enormously contribution to get a much higher $\epsilon$. This offers the \textcolor{black}{possibility} of evading the thermal population influence \textcolor{black}{on} enantiomeric-specific state transfer by introducing ro-vibrational transitions in the CTLS.

\subsection{Examples of 1,2-propanediol}

Particularly, we will demonstrate the above idea \textcolor{black}{by using chiral molecules of 1,2-propanediol as an example}. \textcolor{black}{Such molecules {have been considered in the experimental works of enantiomeric-specific state transfer~\cite{DPatteson_exp_PhysRevLett.118.123002} and} enantiodiscrimination~\cite{Patterson_Detection,Patterson_Detection3}.} The rotational constants for 1,2-propanediol molecules are $A/2\pi=8.5244$\,GHz, $B/2\pi=3.6354$\,GHz, and $C/2\pi=2.7887$
GHz \cite{Propanediol_R}. The corresponding three ro-vibrational states are selected as 
$\vert1\rangle=\vert g\rangle\vert J_{\tau M}=0_{00}\rangle$,
$\left\vert 2\right\rangle =\left\vert e\right\rangle \left\vert 1_{01}\right\rangle $,
and $\left\vert 3\right\rangle =\left\vert e\right\rangle \left\vert 1_{10}\right\rangle $~\cite{Y_RealSingleLoop_PhysRevA.98.063401,Koch_Principles_JCP},
where $\vert g\rangle$ and $\left\vert e\right\rangle $ are, respectively, the vibrational lowest and first-excited states  for the motion of OH-stretch with the transition frequency $\omega_{\mathrm{vib}}/2\pi=100.9500$\,THz~\cite{Propanediol_Vib}. Here,
we have adopted the $\left\vert J_{\tau M}\right\rangle $ nomenclature
to designate the rotational states of asymmetric-top molecules. $J$
is the total angular momentum, $\tau$ runs from $-J$ to $J$ in
unit step in order of ascending energy, and $M$ is the magnetic quantum
number \cite{BooK_AngularMomentum} corresponding to the degenerated
magnetic sub-levels.

\begin{figure}[h]
	\centering \includegraphics[scale=0.23]{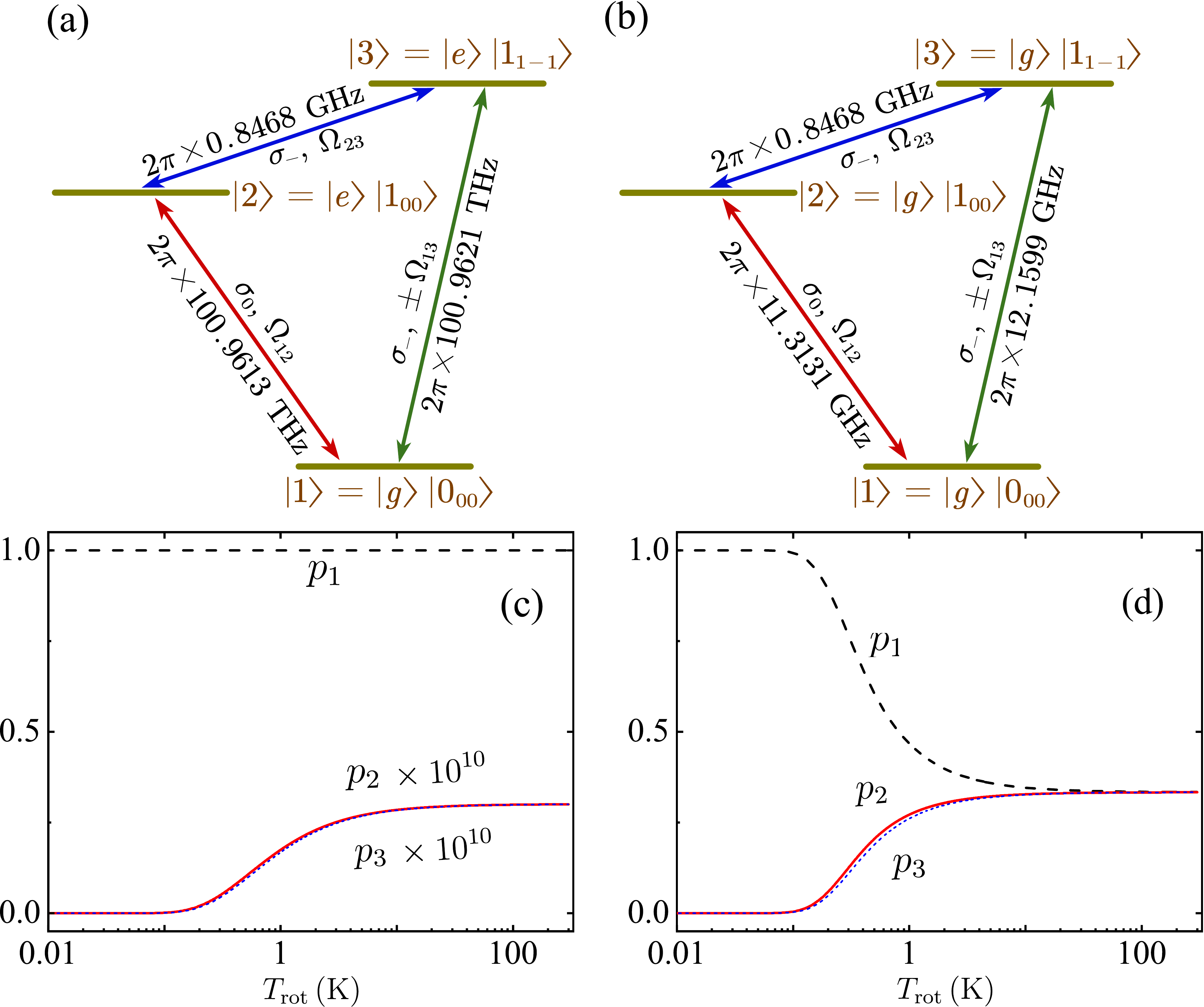} \caption{(Color online) The CTLS of 1,2-propanediol  for (a) the case of ro-vibrational transitions among $\vert1\rangle=\vert g\rangle\vert J_{\tau M}=0_{00}\rangle$, $\vert2\rangle=\vert e\rangle\vert1_{01}\rangle$, and $\vert3\rangle=\vert e\rangle\vert1_{10}\rangle$; (b) the case of purely rotational transitions among  $\vert1\rangle=\vert g\rangle\vert 0_{00}\rangle$, $\vert2\rangle=\vert g\rangle\vert1_{01}\rangle$, and $\vert3\rangle=\vert g\rangle\vert1_{10}\rangle$.
		Here,  $\vert g\rangle$ and $\left\vert e\right\rangle $ are, respectively, the vibrational lowest and first-excited states  for the vibrational motion of OH-stretch with  the corresponding  vibrational transition frequency $\omega_{\mathrm{vib}}/2\pi=100.9500$\,THz. By choosing three special polarized electromagnetic fields, the real single loop CTLS come true~\cite{Y_RealSingleLoop_PhysRevA.98.063401,Koch_Principles_JCP}. Panels (c) and (d) show the thermal populations of the three levels in the CTLS versus the effective rotational temperature with the effective vibrational temperature $T_{\mathrm{vib}}=300$\,K according to the cases (a) and (b), respectively. \label{fig:Structure_Pop}}
\end{figure}

Figure~\ref{fig:Structure_Pop}(a) displays the CTLS
of 1,2-propanediol for the case of ro-vibrational transitions.
The corresponding initial thermal
population with different rotational temperature $ T_\mathrm{rot} $ for each selected state of the CTLS is numerically shown
in Fig.~\ref{fig:Structure_Pop}(c) at the effective vibrational temperature $T_{\mathrm{vib}}=300$\,K according to Eq.~(\ref{eq:Boltzmann_Distribution}).
As the vibrational transition frequency of OH-stretch is about $2\pi\times100.9500$\,THz, there will be little initial thermal occupation for the excited states even at high rotational temperature (e.g. $ T_{\mathrm{rot}}=300 $\,K). In other words, among the selected three states $ \ket{1} $, $ \ket{2} $, and $ \ket{3} $ in the CTLS, we can assume that \textcolor{black}{initially} only the ground state $ \ket{1} $ is thermally populated.

As a comparison, we consider the case \textcolor{black}{of purely rotational transitions in the CTLS }with the same vibrational state $\left\vert \psi_{\mathrm{vib}}\right\rangle =\left\vert g\right\rangle $. 
The rotational transitions among $\vert1\rangle=\vert g\rangle\vert  0_{00}\rangle$, $\vert2\rangle=\vert g\rangle\vert1_{01}\rangle$, and $\vert3\rangle=\vert g\rangle\vert1_{10}\rangle$ \textcolor{black}{of the CTLS are displayed by Fig.~\ref{fig:Structure_Pop}(b), and the corresponding thermal populations are shown in Fig.~\ref{fig:Structure_Pop}(d).}
As expected, the excited states $\left\vert 2\right\rangle =\vert g\rangle\vert1_{01}\rangle$
and $\left\vert 3\right\rangle =\vert g\rangle\vert1_{10}\rangle$ have a considerable population at the  typical rotational temperature $T_{\mathrm{rot}}= 10$\,K in experiments.



\begin{figure}[H]
	\centering \includegraphics[scale=0.3]{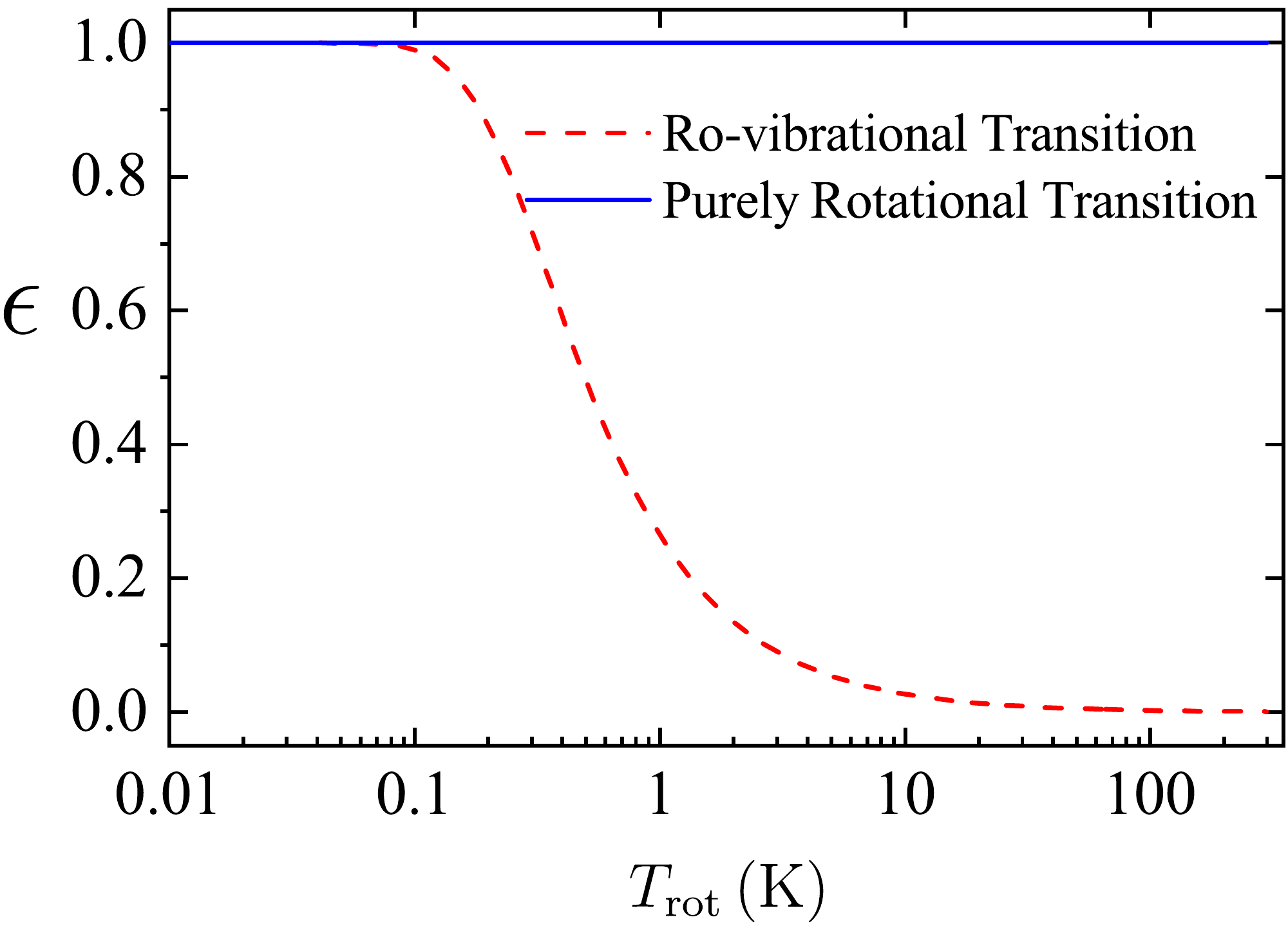} \caption{(Color online) The enantiomeric excess \textcolor{black}{for the case of ro-vibrational transitions {(red dashed line)} or purely rotational transitions {(blue solid line)}} as a function of rotational temperature $T_{\mathrm{rot}}$ with vibrational temperature
		$T_{\mathrm{vib}}=300$\,K. \textcolor{black}{ The other parameters} are the same as those in
		Fig.~\ref{fig:Structure_Pop}. \label{fig:Enantiomeric_Excess}}
\end{figure}


For the racemic mixture of chiral molecules, the initial thermal populations \textcolor{black}{in the states $ \ket{n}_L $ and $ \ket{n}_R $ will be the same}. According to the method of the dynamic operation of ultrashort pulses as discussed in Sec.~\ref{ro-vibrationalStates_SSET}, the enantiomeric-specific state transfer can be realized. Specifically, we give the enantiomeric excess $\epsilon$ to depict the obtained state-specific enantiomeric enrichment. The enantiomeric excess $\epsilon$ as a function of rotational temperature $T_{\mathrm{rot}}$ with vibrational temperature $ T_{\mathrm{vib}}=300 $\,K is shown in Fig.~\ref{fig:Enantiomeric_Excess}. {It shows the enantiomeric excess $\epsilon$ is nearly $100\%$ (see the blue solid line in Fig.~\ref{fig:Enantiomeric_Excess}) for the case of ro-vibrational transitions \textcolor{black} with the effective rotational temperature $ T_\mathrm{rot} $ in a wide temperature range, even at $ 300 $\,K. However, for the case of purely rotational transitions, $\epsilon$ will rapidly decrease when $ T_{\mathrm{rot}}\gtrsim 0.1 $\,K (see the red dashed line in Fig.~\ref{fig:Enantiomeric_Excess}). For instance, at $T_{\mathrm{rot}}=10$\,K, which is near to} the cooled rotational temperature achieved in {the} experiment {for the chiral molecules of 1,2-propanediol~\cite{DPatteson_exp_PhysRevLett.118.123002}}, $\epsilon$ is only {about $2\%$}.

\subsection{{Ratio of pure enantiomers to the chiral mixture}}

Based on the chiral-molecule CTLS composed of ro-vibrational transitions instead of purely rotational ones, \textcolor{black}{we have shown above that} the thermal population influence \textcolor{black}{on} enantiomeric-specific state transfer is negligible. In the realistic case, \textcolor{black}{besides the selected three states (i.e., $\vert1\rangle=\vert g\rangle\vert 0_{00}\rangle$, $\left\vert 2\right\rangle =\left\vert e\right\rangle \left\vert 1_{01}\right\rangle $, and $\left\vert 3\right\rangle =\left\vert e\right\rangle \left\vert 1_{10}\right\rangle $) in the desired CTLS}, the other states out of the CTLS will also be thermally occupied. For the left- or right-handed molecules, we can introduce $P_{n}$ ($ n=1,2,3 $) to describe the proportion of state $\vert n\rangle$ to the related total population as
\begin{equation}
P_{n}=\frac{1}{Z_{\mathrm{tot}}}\mathcal{P}_{n,\mathrm{vib}}\mathcal{P}_{n,\mathrm{rot}},
\end{equation}
\textcolor{black}{where $ Z_{\mathrm{tot}}=\sum_{j}\mathcal{P}_{j,\mathrm{vib}}\mathcal{P}_{j,\mathrm{rot}} $ is the total partition function \textcolor{black}{with $j$ summing for all the states $\ket{v}\ket{J_{\tau_M}}$.} Here $\ket{v}$ represents the vibrational ground state $\ket{g}$, vibrational first excited state $\ket{e}$, and vibrational higher-energy excited ones. }

\begin{figure}[H]
		\centering{}\includegraphics[scale=0.3]{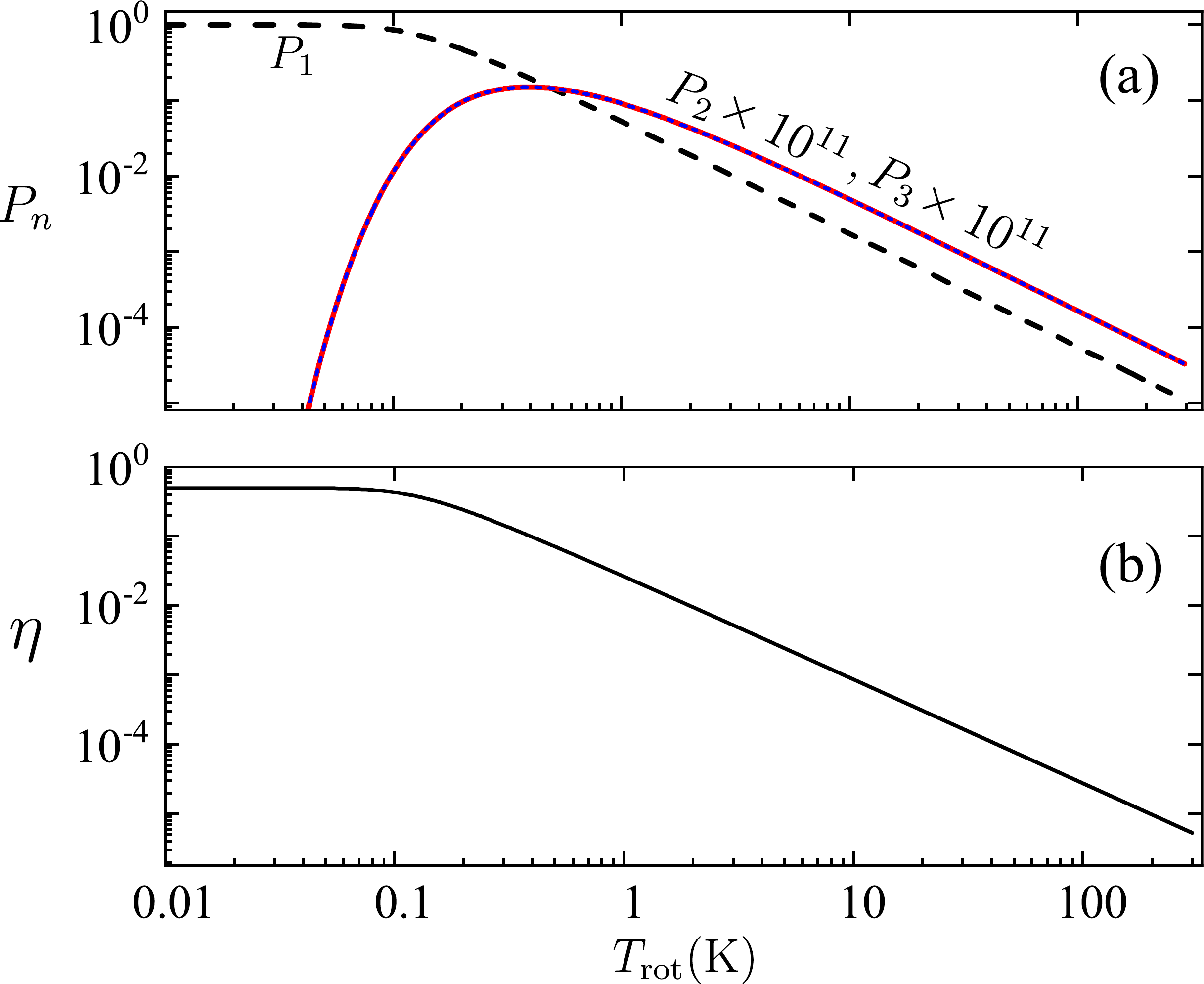}\caption{(Color online) (a) The proportion for the selected three states $\vert1\rangle$ {(black dashed line), $\vert2\rangle$ (red solid line), and $\vert3\rangle$ (blue dotted line) when all of the other vibrational and rotational states out of the CTLS are considered}; (b) The \textcolor{black}{ratio of pure enantiomers}  to the racemic mixture at different rotational temperature $T_{\mathrm{rot}}$. The other parameters are the same as those in Fig.~\ref{fig:Structure_Pop}.
			\label{fig:PY-Times}}
	\end{figure}

 \textcolor{black}{The proportions $ P_n $  of the selected ro-vibrational states $\vert 1 \rangle$, $\vert 2 \rangle$, and $\vert 3 \rangle$  are shown in Fig.~\ref{fig:PY-Times}(a). When $  T_\mathrm{rot}\le 0.1$\,K, $ P_1 $ approximates to 1 and $ P_{2,3} $ approximates to 0, which means most molecules are in the ground state of the CTLS. {Then, after a single operation of enantiomeric-specific state transfer as given in Sec.~\ref{ro-vibrationalStates_SSET}, most  right-handed molecules will occupy  $ \ket{2}_R $, while the left-handed molecules will be approximately  not populated  in $ \ket{2}_L $. Moreover, via the energy-dependent spatially separated processes~\cite{Shapiro_OpticalSwitch_PhysRevLett.90.033001,Shapiro_PopulationTransfer_PhysRevLett.87.183002,Vitan_Shortcut_PhysRevLett.122.173202}, the pure enantiomers with nearly all the right-handed molecules can be obtained from the chiral mixture. When $ T_\mathrm{rot}> 0.1$\,K, the total proportions in the CTLS  $ (P_1+ P_2 +P_3 \simeq P_1 $) decrease rapidly with the increase of the temperature. For an example, at $ T_\mathrm{rot}=10 $\,K, $ P_1 \simeq 0.1\% $ ($ P_2+P_3\simeq0 $). Then, only 0.1\% pure enantiomers with right-handed molecules  can be  obtained.} }
	
 \textcolor{black}{It is clearly that the ratio of the obtained pure enantiomers to the chiral mixture, $\eta$, is determined by the initial population in $ \ket{1}_R $. Specifically, we have}
\begin{equation}
	\mathcal{\eta}=\frac{1}{2}P_1,
\end{equation}
where the factor $ 1/2 $ results from the fact that the racemic mixture consists of equal amounts of left- and right-handed molecules.  \textcolor{black}{Accordingly, we show {in Fig.~\ref{fig:PY-Times}(b) the ratio $ \eta $  at different rotational temperature $T_{\mathrm{rot}}$.} When the  rotational temperature $T_{\mathrm{rot}}> $0.1\,K,  the ratio $ \eta $ will decrease rapidly. In order to get more pure enantiomers from the chiral mixture, the rotational temperature should be further lowered and/or the \textcolor{black}{processes }of the enantiomeric-specific state transfer as well as the subsequent spatial separation should be repeated.}

\section{Conclusion\label{Conclution}}

In conclusion, we have theoretically explored the thermal population influence \textcolor{black}{on} the enantiomeric-specific state transfer based on the CTLS of chiral molecules by considering the ro-vibrational transitions rather than purely rotational transitions. Correspondingly, two infrared pulses and one microwave pulse  are used to respectively couple with the ro-vibrational and rotational transitions. {This scheme was first proposed for enantiomeric-specific state transfer in Ref.~\cite{Koch_Principles_JCP}. Besides the promising features of this scheme emphasized in Ref.~\cite{Koch_Principles_JCP},} we have thoroughly discussed its advantage in evading thermal population influence \textcolor{black}{on the} enantiomeric-specific state transfer over {the three-microwave-pulses  scheme} by coupling purely rotational transitions with the example of 1,2-propanediol. The adverse {influence} of thermal population in \textcolor{black}{other methods of } enantioseparation~\cite{Jia_2010_OpticalPause,Effective_two_level_Model,
Vitan_Shortcut_PhysRevLett.122.173202,YLi_SpatialSepatation_PhysRevLett.99.130403,
LiXuan_SpatialSeparation,Koch_Principles_JCP,QuantumControl} and enantiodiscrimination~\cite{WZJ_Detection,Patterson_Detection,DPatterson_Detection2,
Patterson_Detection3,YY_Chen,CY_ACStark,Lobsiger,Shubert,Delta} based on the CTLS can also be evaded by adopting our similar idea.

Usually, the ro-vibrational transition momenta corresponding to infrared pulses are typically smaller than the rotational transition momenta corresponding to microwave pulses. Since the intensity of experimental available \textcolor{black}{infrared} pulse will be much stronger than that of microwave pulse, the magnitude of the coupling strengths of these three transitions in our scheme can be comparable~\cite{Koch_Principles_JCP,Diastereomer}. \textcolor{black}{It is worth noting that {the operation time of enantiomeric-specific state transfer based on the CTLS} can be about $ 700 $~ns in experiments~\cite{DPatteson_exp_PhysRevLett.118.123002}. {That means such a method} is implementable comparing with the relaxation time $ 6~\mu $s~\cite{DPatteson_exp_PhysRevLett.118.123002}}. In addition, the precise control of relative phases of the three fields plays an important role for enantiomeric-specific {state} transfer, which has been realized experimentally in the three-microwave-pulses scheme~\cite{DPatteson_exp_PhysRevLett.118.123002,Exp_Coherent_Enantiomer_Selective,	Exp_supersonic,Exp_Cooling_Supersonic}. As for the scheme {by considering the ro-vibrational transitions}, it had been theoretically proposed in Ref.~\cite{Koch_Principles_JCP} that the precise control of relative phases of the three fields may be realized by phase-locking two infrared pulses to a common reference standard (e.g., frequency comb) controlled by a microwave reference pulse.

Moreover, the polarizations of the three \textcolor{black}{electromagnetic }pulses can not \textcolor{black}{be} in the same plane when the molecular rotations are considered~\cite{Y_RealSingleLoop_PhysRevA.98.063401,Koch_Principles_JCP}. This will bring the phase-matching problem in realistic cases and eventually reduce the achieved enantiomeric enrichment greatly~\cite{KKLehmann2}. In experiments~\cite{DPatteson_exp_PhysRevLett.118.123002,Exp_Coherent_Enantiomer_Selective,
Exp_supersonic,Exp_Cooling_Supersonic}, phase-matching condition is approximately satisfied,  when  the length scale of the practical excitation volumes is much smaller than the largest wavelength of three applied fields driving the chiral molecules~\cite{KKLehmann2}. Considering this, {our} scheme  with the largest wavelength in the micreowave region has the advantage over the three-infrared-pulses scheme which  can be also used to evade thermal population influence in enantiomeric-specific state transfer.

\section*{ACKNOWLEDGMENTS}

This work was supported by the National Key R\&D Program of China
grant (2016YFA0301200), the Science Challenge Project (under Grant No. TZ2018003) and the Natural Science Foundation of China (under
Grants No. 11774024, No. 11534002, No. U1930402, and  No. 11947206).

\bibliographystyle{apsrev4-1}

\end{document}